\documentclass[a4]{article}
\usepackage{wospa-spconf,amsmath,epsfig}
\usepackage{subfigure}
\usepackage{url}
\def\trans{\mbox{\tiny $T$}}

\title{PUPIL LOCALIZATION AND TRACKING FOR VIDEO-BASED IRIS BIOMETRICS}

\name {N. Benletaief$^1$, A. Benazza-Benyahia$^1$ and S. Derrode$^2$}
\address{
 \begin{minipage}[b]{0.33\textwidth}
  \centering
 \small{$^1$ URISA, SUP'COM, Tunisia\\
 \url{benletaief.nedra@gmail.com}\\ \url{benazza.amel@supcom.rnu.tn}
}
\end{minipage}
 \begin{minipage}[b]{0.33\textwidth}
 \centering
 \small{$^2$ Institut Fresnel (CNRS UMR 6133),\\
  Ecole Centrale de Marseille, France\\
 \url{stephane.derrode@fresnel.fr}
 }
 \end{minipage}
}

\begin{document}
\maketitle
\begin{abstract}
In this paper, we are interested in iris biometric applications. More precisely, our contribution consists in designing both a  pupil detection and a tracking procedure from video sequences acquired by low-cost webcams. The novelty of our approach relies on the fact that it is operational even with a minimal user cooperation and, under bad illuminations and acquisition conditions. A robust classification algorithm is designed to detect the pupil. Moreover,  a  pupil tracker based on the extended Kalman filter is applied in order to reduce the processing time. Experimental results are performed in order to evaluate the performances of the proposed detection and tracking system.
\end{abstract}

\section{Introduction} \label{sec:intro}
Biometry offers new solutions for automatic person recognition applications.  In this respect, much attention is paid to iris recognition since an iris contains rich and unique characteristics of the person which do not alter or evolve with time~\cite{DAUGMAN_06}. The iris identification system starts by localizing in the captured eye image the pupil through which the light enters inside the eye. Hence, the pupil detection is a key issue for achieving satisfactory performances of the whole recognition system. Unfortunately, at this step, many problems could occur. Indeed, the performances are limited if the subject does not strictly cooperate. Besides, the detection step could suffer from suboptimal capture conditions such as specular reflections, occlusions by eyelids and eyelashes.

To handle condition changes, several me\-thods were investigated as the shape-based ones~\cite{shape2} which make use of a prior model of the pupil and the appearance-based ones~\cite{app1} which exploit specific features of  the eye area.  While shape-based me\-thods cannot cope with large shape variations, appearance-based me\-thods are sensitive to light changes, focus and occlusions.

The objective of this paper is to propose a  novel pupil detection and tracking algorithm in a semi-cooperative context, operating under less controlled acquisition conditions. Our idea is to  resort  to video sequences captured by  low cost cameras (typically webcams) in order to select the  best quality images for which the pupil localization procedure ensures the highest   recognition  or identification rates. Furthermore, to meet the requirement of  real time application, we exploit the temporal information through  a fast tracking algorithm which also enhances localization accuracy.

The remainder of this paper is organized as follows. In Section~\ref{sec:iris_localization}, we describe the proposed pupil localization me\-thod. In Section~\ref{sec:iris_tracking}, we show that including a tracking procedure improves the detection step and some conclusions and perspectives are consequently drawn in Section~\ref{sec:Conclusion}.

\section{Proposed pupil localization} \label{sec:iris_localization}
We start with the description of common artifacts, then present related works and finally describe our algorithm.

\subsection{Image artifacts}
Very often, infrared spectrum imaging are employed because the iris strongly reflects infrared light while only the sclera reflects much the visible light. Hence, the pupil corresponds to the darkest region in the image. However, iris images suffer from several artifacts as shown in Fig.~\ref{Fig:prepa}. Indeed, the pupil shape varies with the attitude of the subject relative to the sensor, eyelid closure and individual biometric differences. Moreover, a low contrast alters the shape contour whereas specular reflections from illumination sources corrupt large areas of the region  of interest. Besides, blur could degrade the image: it is due by both the eye motion during exposure time and the defocus when the focal point is outside the depth of field of the object to be captured. Consequently, most of the reported pupil detectors include a preprocessing step to attenuate  such artifacts~\cite{HE_ICIP08}. As far as we are concerned by pupil detection,  we will describe in more detail the preprocessing step we performed in paragraph~\ref{subsubsec:2.2.1}.

\subsection{Related works}
Several methods of pupil localization have been already reported in the literature~\cite{intheeye}. In what follows, we will only focus on the  most employed ones.

The most wide\-spread shape-based approach is due to Daugman~\cite{Daugman}: the objective is to find the inner and outer boundaries of the iris by applying an integro-differential operator. Another alternative was suggested by Wildes~\cite{Wildes}: it consists in computing  an edge map followed by   a  Hough transform in order to detect circles.  Generally, compared to Daugman's method, Wildes's approach is more robust to noise perturbations but at the expense of an increased computational load.  This has motivated extended versions  to reduce the  computational complexity~\cite{Boles,Ma}. Other strategies have also been investigated as those based on the  pupil appearance \cite{app1}. For instance, in~\cite{alex} an iterative thresholding algorithm is considered  to separate between two dark regions that satisfy specific anthropometric constraints using a skin-color model.

The proposed localization algorithm combines the two latest issues to exploit their respective benefits. Indeed, it exploits the  prior information about pupil geometry and intensity distribution and takes into account the processing time as a constraint.

\subsection{Principle of the proposed algorithm}
The goal is to extract the pupil center location. The main steps of the proposed algorithm consist of a morphological preprocessing, a coarse localization and a refined one thanks to an unsupervised classification algorithm. In what follows, we document each of these steps in more detail.\\

\subsubsection {Morphological preprocessing} \label{subsubsec:2.2.1}
\begin{figure}[tb]
    \centering
    \subfigure[Close-up image\label{Fig:prepa}]    {\includegraphics[width=0.23\textwidth]{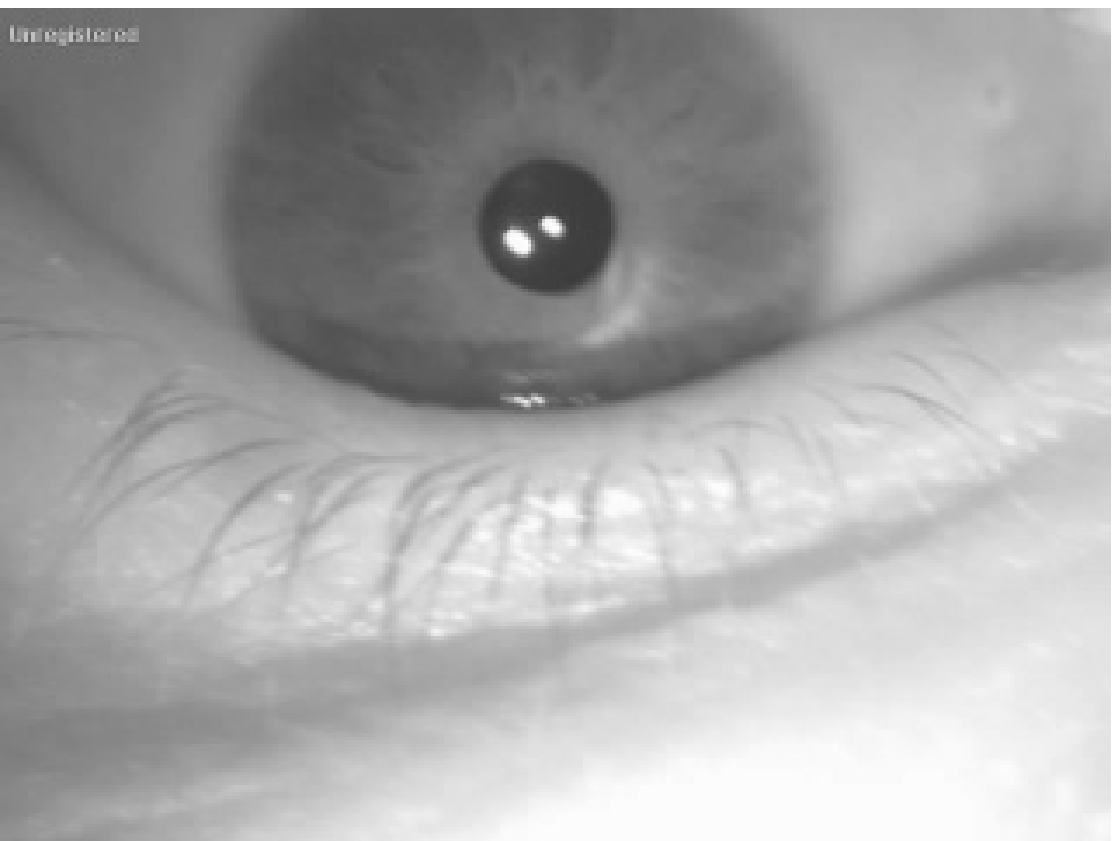}}%
    \subfigure[Preprocessed image\label{Fig:prepb}]{\includegraphics[width=0.23\textwidth]{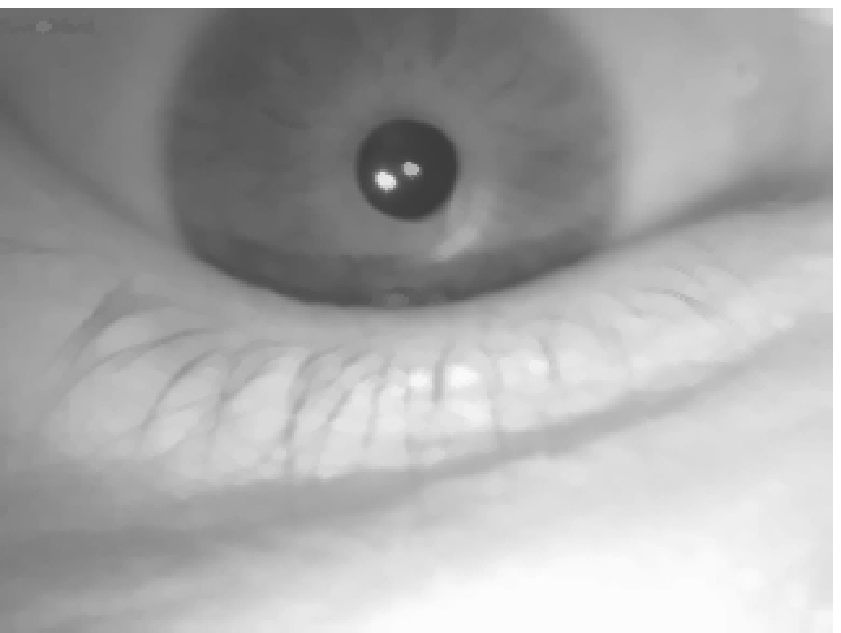}}%
    \caption{Morphological pre-processing.\label{Fig:prep}}
\end{figure}

To reduce the artifacts, we resort to morphological operators as they are able to adapt the processing to the underlying object shape. More precisely, we apply closings for eyelashes impact reduction and openings for white spot and light reflections attenuation. Experimental evaluation using an iris image ($640\times480$ pixels) clearly illustrates the gain from this preprocessing step as shown in Fig.~\ref{Fig:prepb}.\\

\subsubsection {Coarse localization} \label{sec:2.2.2}
\begin{figure}[tb]
    \centering
    \subfigure[Coarse localization\label{Fig:a}]{\includegraphics[width=0.23\textwidth]{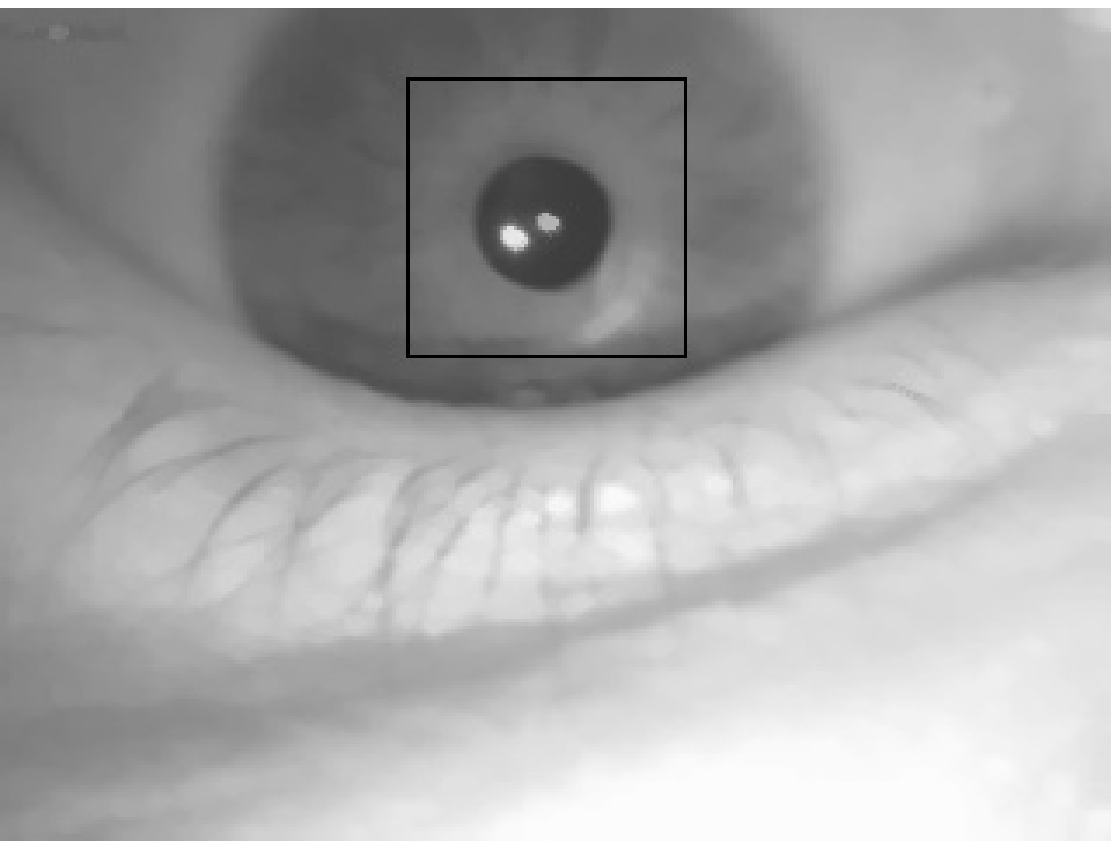}}
    \subfigure[Region of interest\label{Fig:b}] {\raisebox{0.18cm}{\includegraphics[width=0.16\textwidth]{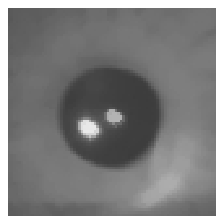}}}
    \caption{Coarse pupil localization.\label{Fig:resultat}}
 \end{figure}

Once the images are enhanced, an initial coarse segmentation is performed to isolate the rectangular region of interest containing the pupil. To this end, a thresholding is applied and  the centroid of the resulting region is considered as a coarse position of the pupil center as illustrated in Fig.~\ref{Fig:b}.\\

\subsubsection {Refined localization}\label{sec:2.2.3}
Let $J$ denote the number of pixels falling within the resulting coarse region of interest (CRI). The goal is to classify these pixels having in consideration the fact that the pupil is typically much darker than its surroundings. The refined position of the pupil  center will correspond to the centroid of the class with the minimum average intensity.
In this respect, it is important to consider a meaningful and tractable model of the eye which  accounts for the large variability of eye appearance and dynamics. For each pixel $j=1, \ldots, J$ within the CRI, we consider the feature vector $\mathbf{f}_{j}$ composed by it spatial coordinates $(x_{j},y_{j})$ and its intensity  $I_{j}$. The objective is to classify the set $\mathcal{F}=\left\{ \mathbf{f}_{j} /  j=1,\ldots,J\right\}$ into $C$ classes.

\begin{figure}[tb]
    \centering
    \subfigure[C=5]{\label{Fig:c} \includegraphics[width=0.23 \textwidth]{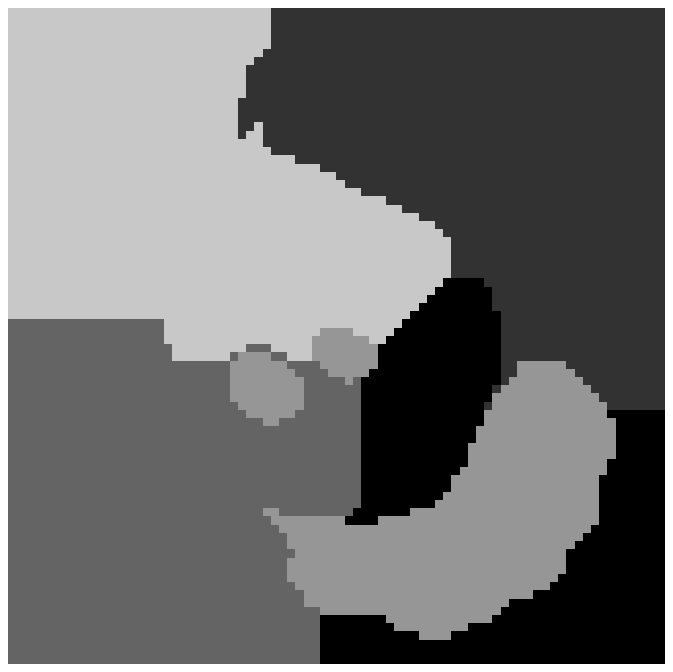}}
    \subfigure[C=2] {\label{Fig:d} \includegraphics[width=0.23 \textwidth]{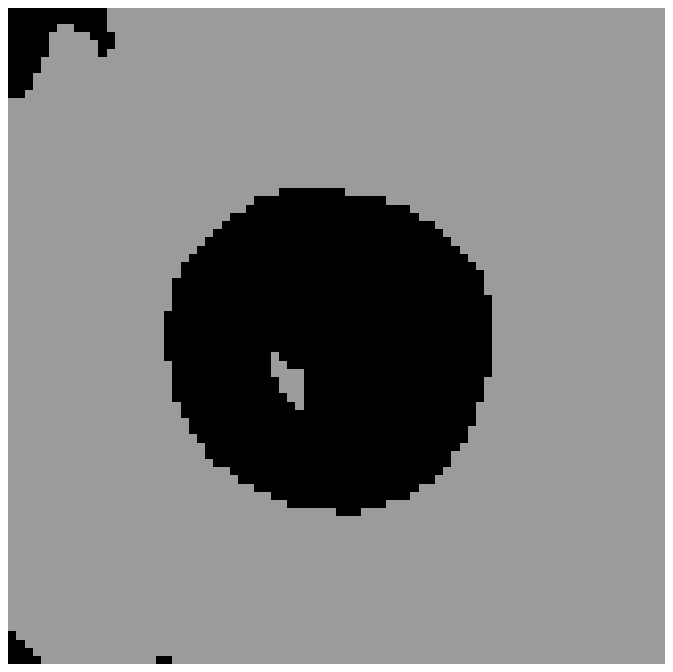}}
    \caption{CAA steps.\label{Fig:resultat}}
 \end{figure}

To this end, the  Competitive Agglomeration algorithm (CAA) is retained as a  unsupervised clustering technique  ~\cite{Frigui}. This choice is motivated by its good classification accuracy  without the need to  specify the number of clusters $C$. It combines the advantages of both partitional and hierarchical classification techniques. Indeed, the CAA begins by assigning the data to a large number of classes, see Fig.~\ref{Fig:c}, then the agglomeration step (which characterizes the hierarchical methods) aims at merging  two or more appropriate clusters, see Fig.~\ref{Fig:d}. More precisely, if $\mathcal{P}=\left(\mathbf{p}_{c}/ c=1,\ldots,C\right)$ represents prototypes of the $C$ clusters, the CAA minimizes the following objective function:
\begin{equation*} \label{eq1}
    \mathcal{J}=\sum_{c=1}^{C}\sum_{j=1}^{J}u_{cj}^{2}\parallel \mathbf{\mathbf{f}}_{j}-\mathbf{p}_{c}\parallel -\alpha\sum_{c=1}^{C}\left[\sum_{j=1}^{J}u_{cj}\right]^{2}
\end{equation*}
under the constraint $\sum_{c=1}^{C}u_{cj}=1, \;\forall j\in\left\{ 1,\ldots,J\right\}$. Here, $u_{cj}$ is the membership of $\mathbf{\mathbf{f}}_{j}$ to a cluster $c$, $\parallel  \cdot \parallel$ is the Euclidean norm, and $\alpha$ is a positive weight. We have retained the Euclidean norm since the clusters to be detected are circular as near frontal faces and depending on the viewing angle, the pupil appears almost circular. Besides, $\alpha$ aims at ensuring a balance between the two terms of $\mathcal{J}$: the data set should be partitioned into the optimal number of clusters while clusters will be designed to minimize the sum of intra-cluster distances.

\subsection{Results} \label{sec:2.3}
\begin{figure}[tb]
    \centering
    \subfigure[$x$ coordinate]{\includegraphics[width=0.49\columnwidth]{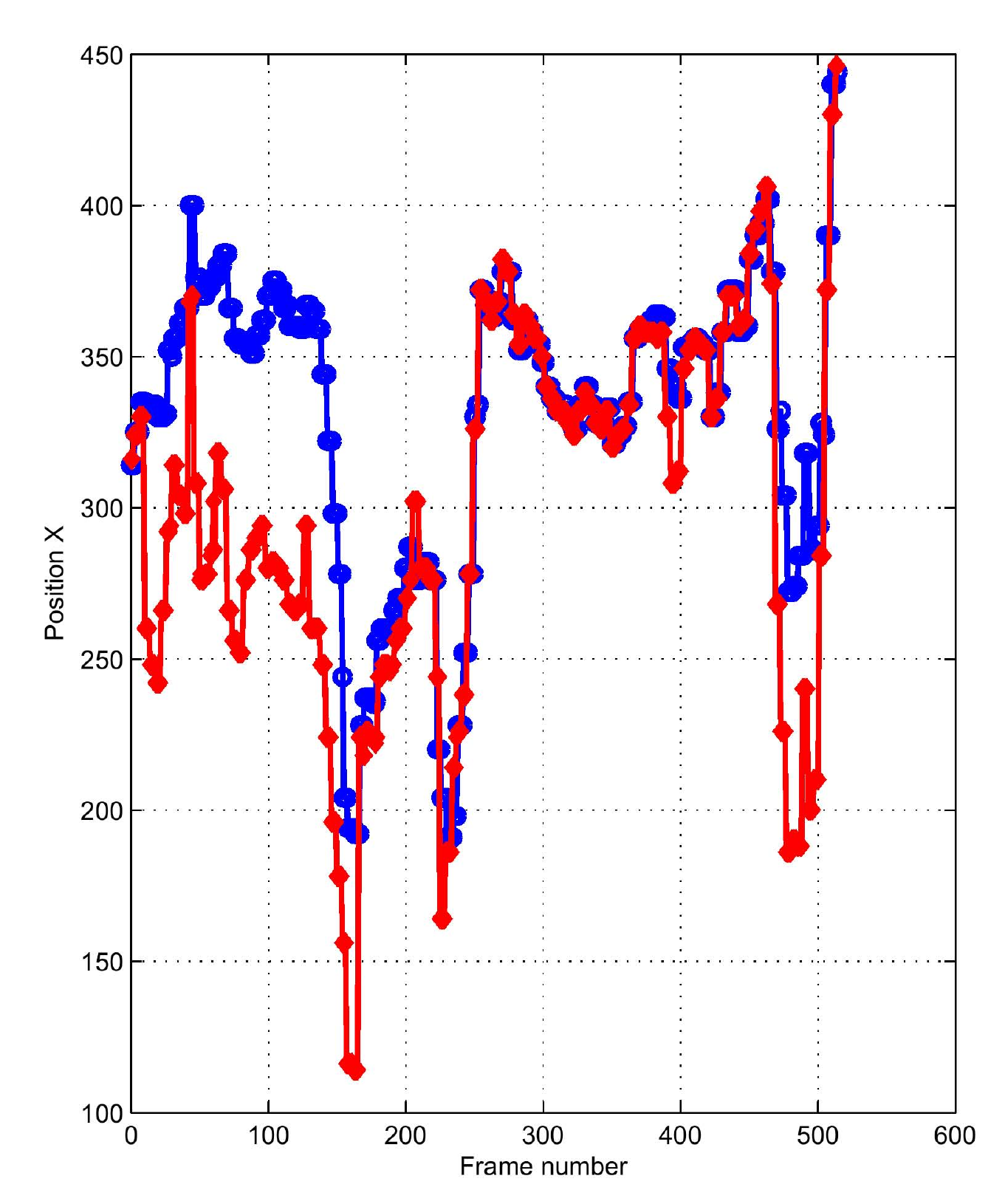}}
    \subfigure[$y$ coordinate]{\includegraphics[width=0.49\columnwidth]{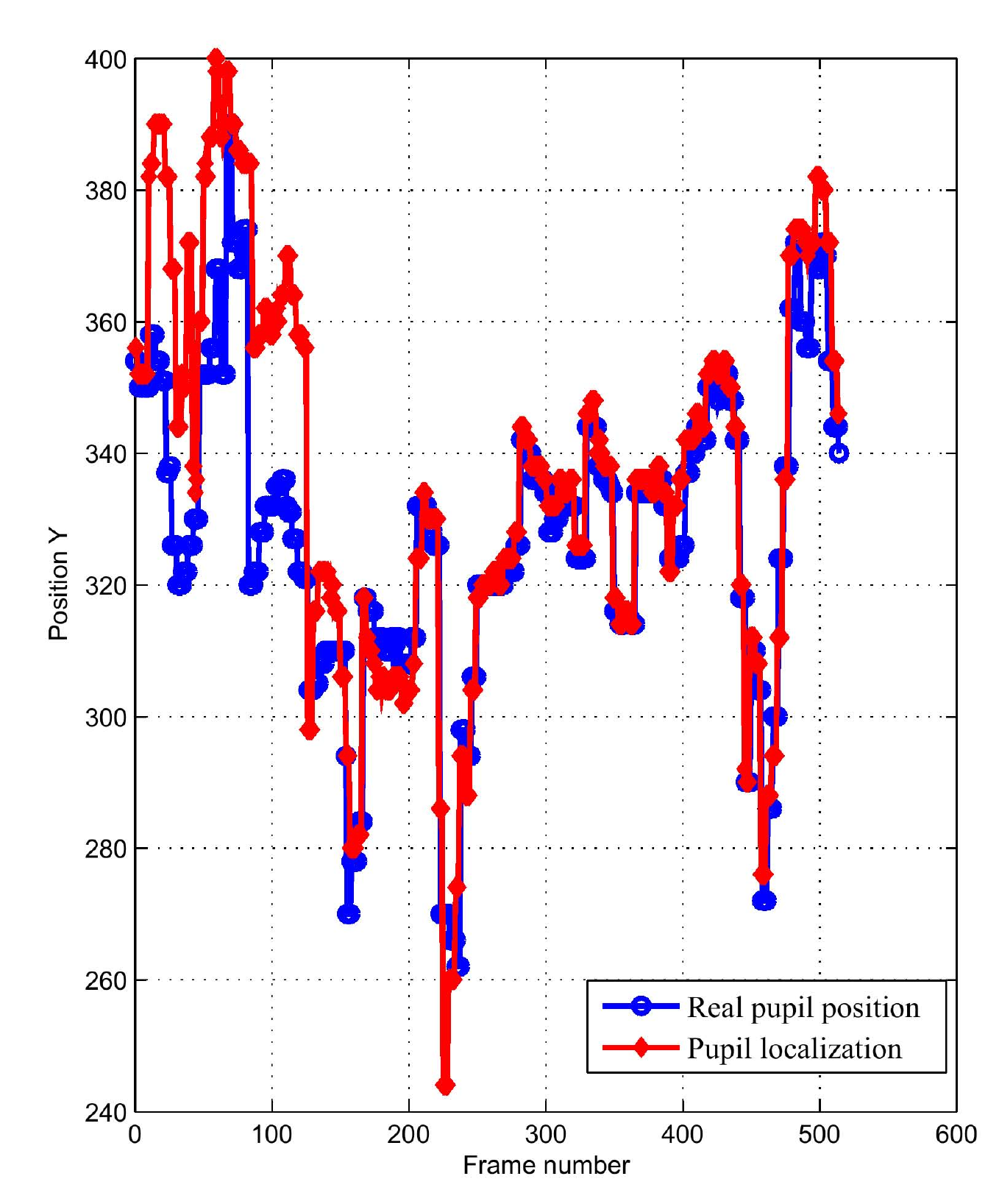}}
    \caption{Localization results of an entire sequence video (no tracking).\label{Fig:localisation}}
\end{figure}

The method has been tested on 517 close-up iris images. The computation time is 172 ms on a single-core CPU (1.73 GHz).  The evolutions of the real positions and the estimated ones along with the time-varying sequence are shown in Fig.~\ref{Fig:localisation}. The proposed localization method is accurate but fails in some situations where images have bad quality. There are several issues including the occlusion of the eye by the eyelids, state of the eye (open/closed), variability in size, reflectivity and head pose. These issues hardly present the same order of magnitude and frequency in the sequence to be drastically eliminated. So, in the next section, we present  a novel solution to tackle such problems.

\section{Pupil tracking} \label{sec:iris_tracking}
The previous detection method separately operates on individual frames and, hence it does not exploit the temporal   information. Consequently, additional refinements in terms of accuracy and computing time should be expected if the pupil localization in the current frame is derived from the estimated position in the previous frames. We start by briefly explain the notations and the theoretic tools used in this aim at the next paragraph. Then, we conclude from the experiments.

\subsection{Problem statement}
 In other words, it is suggested to apply a tracking method based on the concept of  Kalman filtering (EKF)~\cite{kf}. In our work, we propose to resort to the extended Kalman filter in order to take into account the nonlinearity of the observation equation. More precisely, the state vector $\mathbf{s}_{k}$ of the considered system in the $k$-frame is defined by:
\begin{equation*}
    \mathbf{s}_{k}=\left(x_{k},dx_{k},y_{k},dy_{k}\right)^{\trans}
\end{equation*}
where $(x_{k},y_{k})$ is the pupil position and $\left(dx_{k},dy_{k}\right)$ is the pupil displacement between $(k-1)T$ and $kT$,  $T$ being the temporal sampling period. The dynamics of the pupil is modeled as a first order auto-regressive process involving a Gaussian additive noise depicted by the following linear dynamic state equation:
\begin{equation*} 
    \mathbf{s}_{k}=\mathbf{A}\mathbf{s}_{k-1}+\mathbf{q}_{k}
\end{equation*}
where $\mathbf{q}_{k}$ is a realization of a zero-mean white Gaussian process $\mathcal{N}(\mathbf{0},\mathbf{Q})$ and
\begin{equation*}
    \mathbf{A}=\left[
        \begin{array}{cccc}
            1 & T & 0 & 0\\
            0 & 1 & 0 & 0\\
            0 & 0 & 1 & T\\
            0 & 0 & 0 & 1
        \end{array}
        \right].
\end{equation*}
In our experiments, we have noted that the measurements $\mathbf{c}_{k}=(\hat{x}_{k},\hat{y}_{k})^{\trans}$ -- which are the estimated position $(\hat{x}_{k},\hat{y}_{k})$ resulting from the  proposed localization method -- are  related to the state vector according to a nonlinear equation:
\begin{equation*}
    \mathbf{c}_{k}=\mathbf{h}_{k}(\mathbf{s}_{k})+\mathbf{r}_{k}
\end{equation*}
where $\mathbf{r}_{k}$ is the realization of the observation noise assumed to be $\mathcal{N}(\mathbf{0},\mathbf{R})$, $\mathbf{r}_{k}$  and $\mathbf{q}_{k}$ are mutually independent and $\mathbf{h}_{k}$ is a nonlinear function.\\
After several tests, we success to formulate an approximate and closed-form expression of this function  that minimizes the mean square error between the real observations and those given by the model:
        \begin{equation*}
            \mathbf{h}_{k}(\mathbf{s}_{k})=e^{-b(\mathbf{H}\mathbf{s}_{k}-\mathbf{c}_{k-1})} \: \mathbf{H} \; \mathbf{s}_{k},
        \end{equation*}
        where $b$ is a parameter and
        \begin{equation*}
            \mathbf{H}=\left(\begin{array}{cccc}
                1 & 0 & 0 & 0\\
                0 & 0 & 1 & 0\end{array}\right).
        \end{equation*}
        We assume that $e^{(.)}$ for a vector is the vector consisting of the exponential of its components.
It is worth noting that $b$ was estimated according to the least mean squares. The EKF aims at estimating the state $\mathbf{s}_k$ in the sense of a mean square error from the current and past observations.

\subsubsection {Extended Kalman Filter}\label{sec:3.2.2}
In this respect, the EKF only uses the first order terms in the Taylor series expansion of the nonlinear equation. Given the  estimated state $\hat{\mathbf{s}}_{k/k}$ at the $k$  given all the measurement up to   $k$  with the corresponding estimation covariance matrix  $\mathbf{P}_{k/k}$, the EKF algorithm consists of the following steps.
\begin{enumerate}
    \item Initialize filter at $k=0$
    \item The state prediction at time instant $k$ and the corresponding covariance matrix are:
        \begin{eqnarray*}
            &\hat{\mathbf{s}}_{k/k-1} &= \mathbf{A}\hat{\mathbf{s}}_{k-1/k-1}\\
            &\mathbf{P}_{k/k-1}       &= \mathbf{A}\mathbf{P}_{k-1/k-1}\mathbf{A}^{\trans}+\mathbf{Q}.
        \end{eqnarray*}
    \item The Kalman gain is:
        \begin{equation*}
            \mathbf{K}_{k}=\mathbf{P}_{k/k-1}\mathbf{H}_{k}^{\trans}\left[\mathbf{H}_{k}\mathbf{P}_{k/k-1}\mathbf{H}_{k}^{\trans}+\mathbf{R}\right]^{-1}
        \end{equation*}
        where $\mathbf{H}_{k}$ is the Jacobian of the nonlinear function calculated at the state prediction:
        \begin{equation*}
            \mathbf{H}_{k}=\left. \frac{\partial\mathbf{h}_{k}}{\partial\mathbf{s^{\trans}}}\right|_{\hat{\mathbf{s}}_{k/k-1}}
        \end{equation*}
    \item The estimation at $k$ given all the measurement up to time instant $k$ and the corresponding covariance matrix are:
        \begin{eqnarray*}
            &\hat{\mathbf{s}}_{k/k} &= \hat{\mathbf{s}}_{k/k-1}+\mathbf{K}_{k}\left[\mathbf{c}_{k}-\mathbf{h}_{k}(\hat{\mathbf{s}}_{k/k-1})\right],\\%
            &\mathbf{P}_{k/k}       &= \left[\mathbf{I}-\mathbf{K}_{k}\mathbf{H}_{k}\right]\mathbf{P}_{k/k-1}.
        \end{eqnarray*}
    \item Increase $k$ to $k+1$ and repeat from step 2.
\end{enumerate}
It is worth noting that the noise covariances should be known \emph{a priori}. In our case, we estimated these covariances by applying the well-known Expec\-tation-Maxi\-mi\-za\-tion algorithm (EM)~\cite{Dempster}.

\subsection{Experimental results}\label{sec:results}
\begin{figure}[tb]
    \centering
    \subfigure[$x$ coordinate]{\includegraphics[width=0.49\columnwidth]{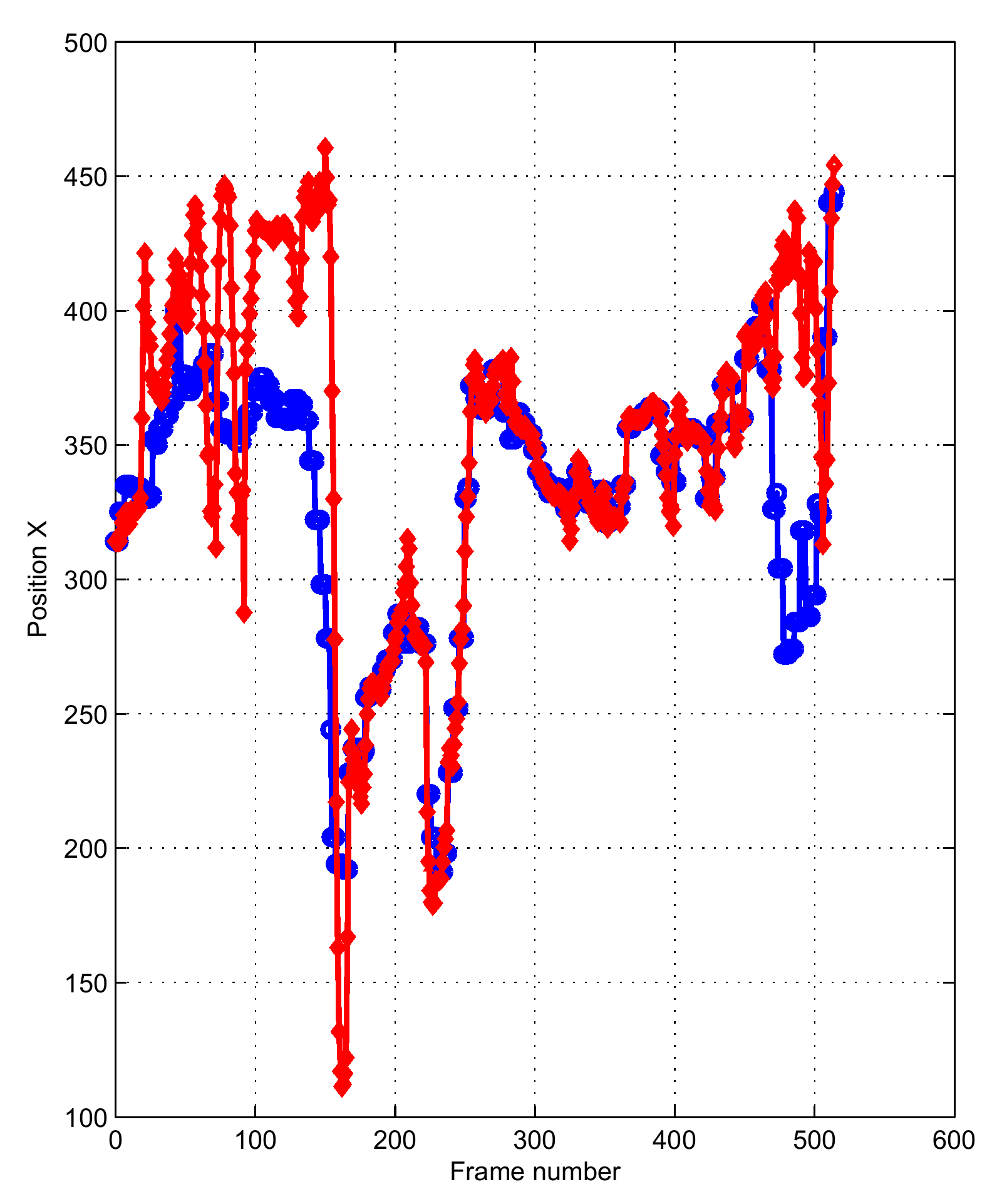}}
    \subfigure[$y$ coordinate]{\includegraphics[width=0.49\columnwidth]{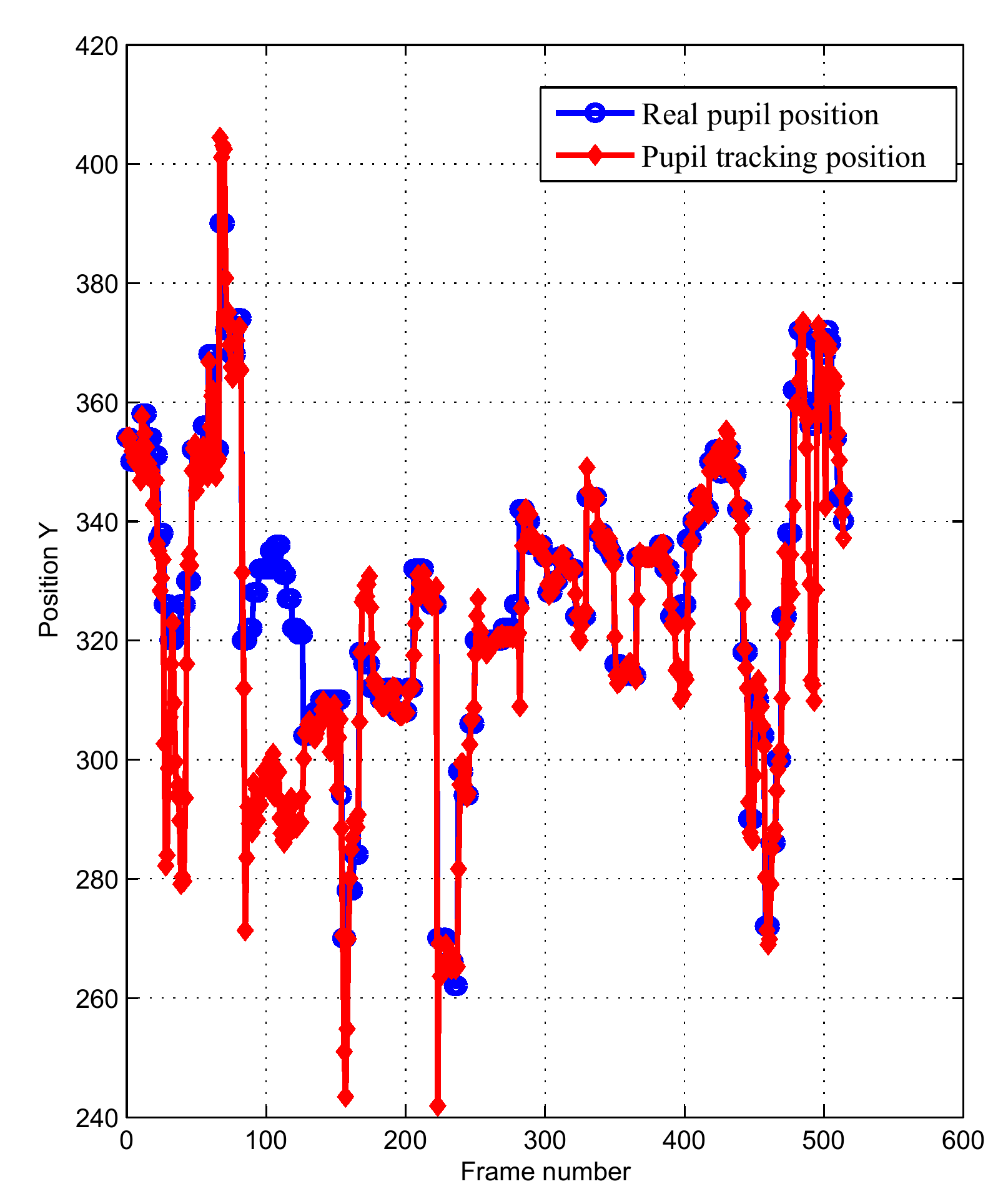}}
    \caption{Localization results of an entire sequence video (with EKF tracking).\label{Fig:Tracking}}
\end{figure}

Fig.~\ref{Fig:Tracking} shows the plots of the estimated position outputted by the EKF-based tracking algorithm. By comparing these plots with those of  Fig.~\ref{Fig:localisation}, it can be noted  that the proposed tracking allows to improve the accuracy of the  pupil localization. Indeed, the achieved average error amounts to 0.217 along the $x$-axis (and to 0.068 along the $y$-axis) for a mere detection and decreases to 0.157 (respectively to 0.034) thanks to the tracking. However,  some errors still remains especially at the moment of backlashes of the eye. These  fast jumps should make null the assumption of the linear state equation and yield to inaccurate estimation of the state vector.
\section{Conclusion} \label{sec:Conclusion}
The video-based eye tracking iris is crucial for biometric applications. But the use of tracking system can profit to many other applications among them medical applications requiring gaze tracking.
There are many related works that are worth further investigating. Firstly, the estimation of covariance matrices could be carried out frame by frame. Furthermore, it seems interesting to address the problem of a suitable nonlinear modeling of the dynamics of the system. Finally, improvement could be expected when some geometric constraints related to the shape of the eye are accounted for.

\end{document}